\documentclass[aps,prd,twocolumn,superscriptaddress,showpacs,preprintnumbers,amsmath,amssymb]{revtex4}

\usepackage{fancyhdr}
\usepackage{lineno}
\usepackage{graphicx}
\usepackage{dcolumn}
\usepackage{bm}
\usepackage{longtable}                                                                                           
        
\usepackage{slashed}

\begin{document}


\title{\boldmath Measurement of the Top Quark Mass Using the Invariant Mass of Lepton Pairs in Soft Muon $b$-tagged Events}
\affiliation{Institute of Physics, Academia Sinica, Taipei, Taiwan 11529, Republic of China} 
\affiliation{Argonne National Laboratory, Argonne, Illinois 60439} 
\affiliation{University of Athens, 157 71 Athens, Greece} 
\affiliation{Institut de Fisica d'Altes Energies, Universitat Autonoma de Barcelona, E-08193, Bellaterra (Barcelona), Spain} 
\affiliation{Baylor University, Waco, Texas  76798} 
\affiliation{Istituto Nazionale di Fisica Nucleare Bologna, $^y$University of Bologna, I-40127 Bologna, Italy} 
\affiliation{Brandeis University, Waltham, Massachusetts 02254} 
\affiliation{University of California, Davis, Davis, California  95616} 
\affiliation{University of California, Los Angeles, Los Angeles, California  90024} 
\affiliation{University of California, San Diego, La Jolla, California  92093} 
\affiliation{University of California, Santa Barbara, Santa Barbara, California 93106} 
\affiliation{Instituto de Fisica de Cantabria, CSIC-University of Cantabria, 39005 Santander, Spain} 
\affiliation{Carnegie Mellon University, Pittsburgh, PA  15213} 
\affiliation{Enrico Fermi Institute, University of Chicago, Chicago, Illinois 60637}
\affiliation{Comenius University, 842 48 Bratislava, Slovakia; Institute of Experimental Physics, 040 01 Kosice, Slovakia} 
\affiliation{Joint Institute for Nuclear Research, RU-141980 Dubna, Russia} 
\affiliation{Duke University, Durham, North Carolina  27708} 
\affiliation{Fermi National Accelerator Laboratory, Batavia, Illinois 60510} 
\affiliation{University of Florida, Gainesville, Florida  32611} 
\affiliation{Laboratori Nazionali di Frascati, Istituto Nazionale di Fisica Nucleare, I-00044 Frascati, Italy} 
\affiliation{University of Geneva, CH-1211 Geneva 4, Switzerland} 
\affiliation{Glasgow University, Glasgow G12 8QQ, United Kingdom} 
\affiliation{Harvard University, Cambridge, Massachusetts 02138} 
\affiliation{Division of High Energy Physics, Department of Physics, University of Helsinki and Helsinki Institute of Physics, FIN-00014, Helsinki, Finland} 
\affiliation{University of Illinois, Urbana, Illinois 61801} 
\affiliation{The Johns Hopkins University, Baltimore, Maryland 21218} 
\affiliation{Institut f\"{u}r Experimentelle Kernphysik, Universit\"{a}t Karlsruhe, 76128 Karlsruhe, Germany} 
\affiliation{Center for High Energy Physics: Kyungpook National University, Daegu 702-701, Korea; Seoul National University, Seoul 151-742, Korea; Sungkyunkwan University, Suwon 440-746, Korea; Korea Institute of Science and Technology Information, Daejeon, 305-806, Korea; Chonnam National University, Gwangju,500-757, Korea; Chonbuk National University, Jeonju 561-756, Korea} 
\affiliation{Ernest Orlando Lawrence Berkeley National Laboratory, Berkeley, California 94720} 
\affiliation{University of Liverpool, Liverpool L69 7ZE, United Kingdom} 
\affiliation{University College London, London WC1E 6BT, United Kingdom} 
\affiliation{Centro de Investigaciones Energeticas Medioambientales y Tecnologicas, E-28040 Madrid, Spain} 
\affiliation{Massachusetts Institute of Technology, Cambridge, Massachusetts  02139} 
\affiliation{Institute of Particle Physics: McGill University, Montr\'{e}al, Qu\'{e}bec, Canada H3A~2T8; Simon Fraser University, Burnaby, British Columbia, Canada V5A~1S6; University of Toronto, Toronto, Ontario, Canada M5S~1A7; and TRIUMF, Vancouver, British Columbia, Canada V6T~2A3} 
\affiliation{University of Michigan, Ann Arbor, Michigan 48109} 
\affiliation{Michigan State University, East Lansing, Michigan  48824}
\affiliation{Institution for Theoretical and Experimental Physics, ITEP, Moscow 117259, Russia} 
\affiliation{University of New Mexico, Albuquerque, New Mexico 87131} 
\affiliation{Northwestern University, Evanston, Illinois  60208} 
\affiliation{The Ohio State University, Columbus, Ohio  43210} 
\affiliation{Okayama University, Okayama 700-8530, Japan} 
\affiliation{Osaka City University, Osaka 588, Japan} 
\affiliation{University of Oxford, Oxford OX1 3RH, United Kingdom} 
\affiliation{Istituto Nazionale di Fisica Nucleare, Sezione di Padova-Trento, $^z$University of Padova, I-35131 Padova, Italy} 
\affiliation{LPNHE, Universite Pierre et Marie Curie/IN2P3-CNRS, UMR7585, Paris, F-75252 France} 
\affiliation{University of Pennsylvania, Philadelphia, Pennsylvania 19104}
\affiliation{Istituto Nazionale di Fisica Nucleare Pisa, $^{aa}$University of Pisa, $^{bb}$University of Siena and $^{cc}$Scuola Normale Superiore, I-56127 Pisa, Italy} 
\affiliation{University of Pittsburgh, Pittsburgh, Pennsylvania 15260} 
\affiliation{Purdue University, West Lafayette, Indiana 47907} 
\affiliation{University of Rochester, Rochester, New York 14627} 
\affiliation{The Rockefeller University, New York, New York 10021} 
\affiliation{Istituto Nazionale di Fisica Nucleare, Sezione di Roma 1, $^{dd}$Sapienza Universit\`{a} di Roma, I-00185 Roma, Italy} 

\affiliation{Rutgers University, Piscataway, New Jersey 08855} 
\affiliation{Texas A\&M University, College Station, Texas 77843} 
\affiliation{Istituto Nazionale di Fisica Nucleare Trieste/Udine, I-34100 Trieste, $^{ee}$University of Trieste/Udine, I-33100 Udine, Italy} 
\affiliation{University of Tsukuba, Tsukuba, Ibaraki 305, Japan} 
\affiliation{Tufts University, Medford, Massachusetts 02155} 
\affiliation{Waseda University, Tokyo 169, Japan} 
\affiliation{Wayne State University, Detroit, Michigan  48201} 
\affiliation{University of Wisconsin, Madison, Wisconsin 53706} 
\affiliation{Yale University, New Haven, Connecticut 06520} 
\author{T.~Aaltonen}
\affiliation{Division of High Energy Physics, Department of Physics, University of Helsinki and Helsinki Institute of Physics, FIN-00014, Helsinki, Finland}
\author{J.~Adelman}
\affiliation{Enrico Fermi Institute, University of Chicago, Chicago, Illinois 60637}
\author{T.~Akimoto}
\affiliation{University of Tsukuba, Tsukuba, Ibaraki 305, Japan}
\author{B.~\'{A}lvarez~Gonz\'{a}lez$^t$}
\affiliation{Instituto de Fisica de Cantabria, CSIC-University of Cantabria, 39005 Santander, Spain}
\author{S.~Amerio$^z$}
\affiliation{Istituto Nazionale di Fisica Nucleare, Sezione di Padova-Trento, $^z$University of Padova, I-35131 Padova, Italy} 

\author{D.~Amidei}
\affiliation{University of Michigan, Ann Arbor, Michigan 48109}
\author{A.~Anastassov}
\affiliation{Northwestern University, Evanston, Illinois  60208}
\author{A.~Annovi}
\affiliation{Laboratori Nazionali di Frascati, Istituto Nazionale di Fisica Nucleare, I-00044 Frascati, Italy}
\author{J.~Antos}
\affiliation{Comenius University, 842 48 Bratislava, Slovakia; Institute of Experimental Physics, 040 01 Kosice, Slovakia}
\author{G.~Apollinari}
\affiliation{Fermi National Accelerator Laboratory, Batavia, Illinois 60510}
% ...

\collaboration{CDF Collaboration\footnote{With visitors from $^a$University of Massachusetts Amherst, Amherst, Massachusetts 01003,
$^b$Universiteit Antwerpen, B-2610 Antwerp, Belgium, 
$^c$University of Bristol, Bristol BS8 1TL, United Kingdom,
$^d$Chinese Academy of Sciences, Beijing 100864, China, 
$^e$Istituto Nazionale di Fisica Nucleare, Sezione di Cagliari, 09042 Monserrato (Cagliari), Italy,
$^f$University of California Irvine, Irvine, CA  92697, 
$^g$University of California Santa Cruz, Santa Cruz, CA  95064, 
$^h$Cornell University, Ithaca, NY  14853, 
$^i$University of Cyprus, Nicosia CY-1678, Cyprus, 
$^j$University College Dublin, Dublin 4, Ireland,
$^k$University of Edinburgh, Edinburgh EH9 3JZ, United Kingdom, 
$^l$University of Fukui, Fukui City, Fukui Prefecture, Japan 910-0017
$^m$Kinki University, Higashi-Osaka City, Japan 577-8502
$^n$Universidad Iberoamericana, Mexico D.F., Mexico,
$^o$University of Iowa, Iowa City, IA  52242,
$^p$Queen Mary, University of London, London, E1 4NS, England,
$^q$University of Manchester, Manchester M13 9PL, England, 
$^r$Nagasaki Institute of Applied Science, Nagasaki, Japan, 
$^s$University of Notre Dame, Notre Dame, IN 46556,
$^t$University de Oviedo, E-33007 Oviedo, Spain, 
$^u$Texas Tech University, Lubbock, TX  79609, 
$^v$IFIC(CSIC-Universitat de Valencia), 46071 Valencia, Spain,
$^w$University of Virginia, Charlottesville, VA  22904,
$^x$Bergische Universit\"at Wuppertal, 42097 Wuppertal, Germany,
$^{ff}$On leave from J.~Stefan Institute, Ljubljana, Slovenia, 
}}
\noaffiliation


\date{\today}
\begin{abstract}         
We present the first measurement of the mass of the top quark in a sample of $t\bar{t}\rightarrow\ell\bar{\nu} b\bar{b} q\bar{q}$ events (where $\ell = e$, $\mu$) selected by identifying jets containing a muon candidate from the semileptonic decay of heavy-flavor hadrons (soft muon $b$-tagging). The $p\bar{p}$ collision data used corresponds to an integrated luminosity of 2~fb$^{-1}$ and was collected by the CDF~II detector at the Fermilab Tevatron. The measurement is based on a novel technique exploiting the invariant mass of a subset of the decay particles, specifically the lepton from the $W$ boson of the $t\rightarrow Wb$ decay, and the muon from a semileptonic $b$ decay. We fit template histograms, derived from simulation of $t\bar{t}$ events and a modeling of the background, to the mass distribution observed in the data and measure a top quark mass of $180.5\pm12.0({\rm stat.})\pm3.6({\rm syst.})~{\rm GeV}/c^2$, consistent with the current world average.
\end{abstract}
\maketitle

A massive top quark plays an important role in the standard model (SM). The mass of the top quark ($m_t$) enters electroweak (EW) precision observables as an input parameter via quantum effects, i.e. loop corrections, and its large numerical value gives rise to sizable corrections that behave as powers of $m_t$ \cite{veltman}. For example, in the theoretical prediction of the $W$ boson mass ($m_W$) within the SM, when these corrections are combined with the logarithmic dependence on the mass of the postulated Higgs boson ($m_H$), a relationship emerges that provides a constraint on $m_H$ from experimental determinations of $m_W$ and $m_t$ \cite{pdg}. Indeed, the strong dependence of the SM radiative corrections  on $m_t$  made it possible to predict the value of $m_t$ \cite{predict} prior to its experimental determination \cite{predictions1, predictions2}. 
Thus, a precision value of $m_t$ is crucial for constraining SM parameters, for high-sensitivity searches for effects of new physics and for stringent consistency tests of models beyond the SM (e.g. supersymmetry). Furthermore, independent measurements of $m_t$ in all final states of $t\bar{t}$ decay provide an important consistency check of the top quark sector of the SM, and might reveal new physics with top-like signatures.

Significant progress has been made recently in reducing the uncertainty in measurements of $m_t$ and in devising alternative and independent techniques. The current best single measurement is determined by reconstructing the full decay chain and computing the invariant mass of the decay products in  $t\bar{t}\rightarrow \ell \bar{\nu} b \bar{b}q\bar{q}$ events, and yields $m_t=172.1\pm1.6$~GeV/$c^2$ \cite{bestmt,bestmt2}. However, this and all the most precise of the current techniques are limited by the common systematic uncertainty in the calorimeter jet energy calibration (jet energy scale, JES). To provide independent measurements, several techniques with minimal dependence on the JES have been proposed. For example, the flight distance of the $b$-hadron from the top decay can be used to infer the mass of the top quark \cite{lxy}, but this method also requires precision track reconstruction to determine the decay length. A proposal has been made \cite{cms} for exploiting the correlation between $m_t$ and the invariant mass of the system composed of a $J/\psi$ (from the decay of a $b$ hadron) and the lepton from the $W$ decay. The advantage is a stronger correlation of this system-mass with $m_t$ than that of individual decay products of the top quark, and thus a better sensitivity to the top quark mass, but the overall branching ratio for this final state is only $\mathcal{O}(10^{-5})$.


We present the first measurement of the mass of the top quark in a sample of $t\bar{t}\rightarrow\ell\bar{\nu} b\bar{b} q\bar{q}$ events (where $\ell = e$, $\mu$) selected by identifying $b$-jets with a candidate muon from semileptonic decay of heavy-flavor hadrons. We have developed a novel technique that exploits the invariant mass of the lepton from the $W$ boson of the $t\rightarrow Wb$ decay, and the muon from a semileptonic $b$ decay. The selection method is complementary to that taking advantage of the long lifetime of $b$-hadrons through the presence of a decay vertex displaced from the primary interaction. Since only $\sim 50\%$ of the sample of $t\bar{t}$ candidates with a semileptonic $b$ decay overlaps the top samples selected by the identification of a displaced vertex, and a still smaller fraction is in common with traditional samples that require all four jets for the mass reconstruction, our technique provides an essentially independent measurement of $m_t$ from these data. Moreover, our observable is largely independent of the JES, because the calorimeter information is used solely for the selection of event candidates, and therefore the result can add a significant amount of information when averaged with those from other measurements.
Including sequential decays of charm, the branching fraction for $b\rightarrow\mu\nu X\simeq 20$\% \cite{pdg} is sizable and since this technique does not require precision secondary vertex reconstruction to suppress backgrounds, it could be an attractive option for the early phase of experiments at the Large Hadron Collider (LHC). Finally, the observable has a higher correlation to the top quark mass than the momentum of the lepton from the $W$ decay alone. A partial reduction in sensitivity will arise from $b$-$W$ mis-pairing, when the lepton from the $W$ decay and the muon from the $b$ semileptonic decay do not originate from the same top quark.

Top quarks are produced at the Tevatron proton-antiproton collider predominantly in pairs of $t$ and $\bar{t}$, and are identified by the SM decay $t\rightarrow Wb$, providing a final state that includes two $W$ bosons and two bottom quarks. $W$'s are identified through their decay to leptons or quarks. Quarks hadronize and are observed as jets of charged and neutral particles. The CDF~II detector is described in detail elsewhere \cite{CDF}. The components relevant to this analysis include the central outer tracker (COT), the central electromagnetic and hadronic calorimeters, the central muon detectors and the luminosity counters. The data sample, produced in $p\bar{p}$ collisions at $\sqrt{s}=1.96~{\rm TeV}$ during Run$~$II of the Fermilab Tevatron, was collected between March 2002 and May 2007 and corresponds to an integrated luminosity of $2.0\pm0.1~{\rm fb}^{-1}$. We select events where one of the $W$ bosons decays to an isolated electron (muon) carrying large transverse energy ($E_T$) (momentum ($p_T$)) \cite{cdfsys} with respect to the beam line, plus a neutrino. We refer to these high-$p_T$ electrons or muons as  primary leptons (PL). The neutrino escapes the detector causing an imbalance of total transverse energy vector, referred to as  missing $E_T$ ($\slashed{E}_T$). The other $W$ boson in the event decays hadronically to a pair of quarks. We take advantage of the semileptonic decay of $B$ hadrons by searching for muons within final-state jets (soft-lepton tagging, or  SLT), in order to identify those jets that result from hadronization of the bottom quarks. 

The event selection starts with an inclusive lepton trigger requiring an electron (muon) with $E_T>18~{\rm GeV}$ ($p_T>18~{\rm GeV}/c$). Further selection requires that candidate electron (muon) PLs are isolated and have $E_T>20$~GeV ($p_T>20$~GeV/$c$) and $|\eta|<1.1$. We define an isolation parameter, $I$, as the calorimeter transverse energy in a cone of opening $\Delta R\equiv \sqrt{(\Delta\eta)^2+(\Delta\phi)^2}=0.4$ around the lepton (not including the lepton energy itself) divided by the electron $E_T$ or muon $p_T$. We select isolated electrons (muons) by requiring $I<0.1$. The event must have $\slashed{E}_T>30$~GeV, consistent with the presence of a neutrino from the $W$ boson decay. Jets are identified using a fixed-cone algorithm with a cone opening of $\Delta R=0.4$ and are constrained to originate from the $p\bar{p}$ collision vertex. Muons inside jets are identified by matching the tracks of the jet, as measured in the COT, with track segments in the muon detectors.
Such a muon with $p_T>3$~GeV/$c$ and within $\Delta R<0.6$ of a jet axis is called an SLT$\mu$ \cite{sltprd}. The probability of misidentifying a hadron as an SLT$\mu$, denoted as the  SLT$\mu$ mistag probability, is measured using a data sample of pions, kaons and protons from $D^*$ and $\Lambda^0$ decays. A Monte Carlo (MC) simulation of $W$+light flavor events is used to model the $\pi$, $K$ and $p$ admixture in light-quark jets. The SLT$\mu$ mistag probability is parametrized as a function of the track $p_T$ and $\eta$, and is seen to describe within $\pm$5\% the number of false SLT$\mu$ tags in light flavor jets of QCD multijet and $\gamma+{\rm jet}$ events.

To reduce background from dimuon resonances and double-semileptonic $B$ hadron decays, we remove events in which the PL muon and SLT$\mu$ are oppositely charged and have an invariant mass consistent with a $Z$, $\Upsilon$ or, irrespectively of the PL flavor, less than 5 ${\rm GeV}/c^2$. We further reject events as candidate radiative Drell-Yan and $Z$ bosons if the tagged jet has an electromagnetic energy fraction above 0.8 and only one track with $p_T>1.0$~GeV/$c$ within a cone of $\Delta R=0.4$ about the jet axis
The jet energies are corrected to account for variations of the detector response in $\eta$ and time, calorimeter gain drifts, non linearity of calorimeter energy response, multiple $p\bar{p}$ interactions in an event and for energy loss in un-instrumented regions \cite{jetcorr}. Finally, the sample is partitioned according to the number of jets with $E_T>20$~GeV and $|\eta|<2.0$  in the event, and at least one jet is required to contain an SLT$\mu$ (defining the $SLT\mu$-tagged $W$+$n$ jets sample). The subset of $W$ plus at least 3 jets is the $t\bar{t}$ candidate sample, and to reduce background from QCD production of $W$ with multiple jets, we additionally require the total transverse scalar energy in the event ($H_T$ \cite{ht}) to be greater than 200~GeV.

Standard model processes that result in the same signature as the $t\bar{t}$ signal are backgrounds to this measurement. There are three dominant backgrounds: the largest one is mistags of $W$+light flavor events, and a smaller contribution is due to $W$ boson in association with heavy flavor jets ($Wb\bar{b}$, $Wc\bar{c}$, $Wc$). Events without $W$ bosons that pass the event selection are typically QCD multijet events where one jet has been reconstructed as a high-$p_T$ lepton, mismeasured jet energies produce apparent $\slashed{E}_T$ and an additional jet contains an SLT$\mu$. A fraction of these events is from $b\overline{b}$ and $c\overline{c}$, where the candidate PL may result from a semileptonic decay of one of the fragmenting heavy quark and the SLT$\mu$ from a semileptonic decay of the other. Other minor backgrounds that can mimic a $W$ boson and an SLT$\mu$ signature include diboson ($WW$, $ZZ$, $WZ$), Drell-Yan$\rightarrow\tau\tau$, single top quark, and residual Drell-Yan$\rightarrow\mu\mu$ events not removed by the dimuon resonance removal. The composition of the data sample used in this analysis has been studied extensively in \cite{sltprd}, where we have measured the production cross section for $p\bar{p}\rightarrow t\bar{t}X$, and is summarized in Table~\ref{tab:sample}. The $W$+jets, QCD multijet and Drell-Yan background are determined using the data, while the remaining backgrounds are estimated from MC simulations. The $W$+1,2 jets samples contain little $t\bar{t}$ events and have a composition similar to the background of the $t\bar{t}$ candidate sample. 
 The simulation of $t\bar{t}$ events is performed using {\sc pythia} \cite{pythia} and {\sc herwig}  \cite{herwig}. The generators are used with the {\sc CTEQ5L} \cite{cteq5} parton distribution functions (PDF). Modeling of $b$ and $c$ hadron decay is provided by {\sc evtgen} \cite{evtgen}. Modeling of  $W+{\rm jets}$ production is performed using {\sc alpgen}  \cite{alpgen}, coupled with {\sc pythia} for the shower evolution and {\sc evtgen} for the heavy flavor hadron decays. Diboson production ($WW$, $ZZ$, $WZ$) and Drell-Yan$\rightarrow\tau\tau$ are determined using {\sc pythia}. Drell-Yan$\rightarrow\mu\mu+{\rm jets}$ events are modeled using {\sc alpgen} while single top production is modeled with {\sc madevent} \cite{madevent}, both with {\sc pythia} showering. The CDF II detector simulation models the response of the detector to particles produced in $p\bar{p}$ collisions. The detector geometry used in the simulation is the same as that used for reconstruction of the collision data. Details of the CDF II simulation, based on the {\sc geant3} package, can be found in \cite{cdfIIsim}.
 

\begin{table}[h]
\caption{\small{Composition of the SLT$\mu$-tagged $W+n$~jets candidate sample \cite{sltprd}. The $H_T>200$~GeV requirement is released for events with fewer than 3 jets.}}
\begin{center}
\begin{tabular}{lccc}
\hline
\hline
Source & $W$+1 jet & $W$+2 jet & $W+\geq3$ jets\\
\hline
$W$+light flavor & 622$\pm$31    & 226$\pm$12       & 52.3$\pm$2.6   \\
$W$+heavy flavor & 145$\pm$55    & 66.6$\pm$25.2    & 14.3$\pm$5.4 \\
QCD multijet   & 91.9$\pm$16.5 & 44.9$\pm$10.4    &  6.9$\pm$1.5 \\
$WW+WZ+ZZ$  	 & 3.8$\pm$0.4 & 7.0$\pm$0.7    & 1.9$\pm$0.3 \\
Drell-Yan$\rightarrow\tau\tau$ & 2.6$\pm$0.6 & 1.5$\pm$0.4 & 0.6$\pm$0.3 \\
Drell-Yan$\rightarrow\mu\mu$  &  6.0$\pm$1.2 & 4.1$\pm$0.9 & 0.8$\pm$0.5 \\
Single top  	& 4.4$\pm$0.4 & 9.0$\pm$0.7 & 2.7$\pm$0.2 \\
\hline
Total background & 876$\pm$54 & 359$\pm$24 & 79.5$\pm$5.3 \\
$t\bar{t}$ ($\sigma_{t\bar{t}}=9.1~{\rm pb}$) & 3.5$\pm$0.2 & 31.8$\pm$1.0 & 168.5$\pm$5.3 \\
Data  &  892 & 384 & 248   \\
\hline
\hline
\end{tabular}
\end{center}
\label{tab:sample}                 
\end{table}

\begin{figure}[h]
\begin{center}
\includegraphics[width=6.5cm]{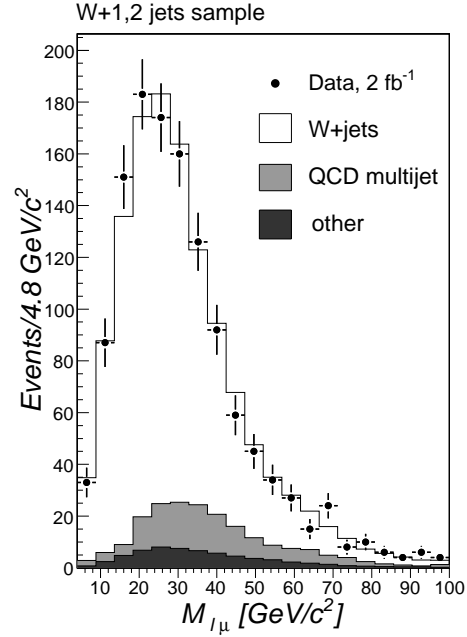} 
\end{center}
\caption{{\small The predicted and observed $M_{\ell \mu}$ distributions in the sample of $W$+1,2 jet SLT$\mu$-tagged events. The predicted distributions are stacked.}}
\label{fig:Control_IM_12Ja}
\end{figure}

We compute the invariant mass ($M_{\ell \mu}$) between the PL and the SLT$\mu$ in the $t\bar{t}$ candidates sample. In rare cases where there is more than one SLT$\mu$ tag in the same jet, or more than one  SLT$\mu$ tagged jet in the same event, we use the SLT$\mu$ candidate that has the best match between the COT track and the track segment in the muon detectors. No attempt is made to choose the correct pairing from the decay chain of the two top-quarks. The electric charge of the SLT$\mu$ for instance is not an effective flavor selector due to abundant sequential $b\rightarrow c\rightarrow\mu$ decays. When the wrong pairing is chosen, there is still sensitivity to the top quark mass  due to the boost of the SLT$\mu$ and the PL. The distribution of $M_{\ell \mu}$ is given by the contribution of $t\bar{t}$ and background events. For the background, the $M_{\ell \mu}$ distribution of QCD multijet events is derived from the data themselves in the kinematic-region of $I>0.15$, $\slashed{E}_T>30$~GeV, topologically close to the signal region, while for other background sources we use MC simulation. We check the background model in $W$+1,2 jet SLT$\mu$-tagged data events, a sample with a similar composition as the background to $t\bar{t}$ candidates. We find the predicted and observed distributions of $M_{\ell \mu}$ (Figure \ref{fig:Control_IM_12Ja}) to be in agreement with a $p$-value of 55\%, as given by the Kolmogorov-Smirnov test.

\begin{figure}[h]
\begin{center}
\includegraphics[width=6.5cm]{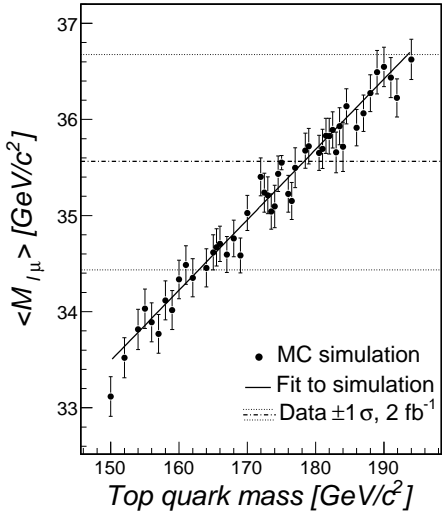}
\end{center}
\caption{{\small The correlation between the mean value of the $M_{\ell \mu}$ histograms from simulated $t\bar{t}$ and background samples, and the input $m_t$. The continuous line shows a linear fit to the points.}}
\label{fig:Control_IM_12Jb}
\end{figure}  

We construct a set of template histograms of the $M_{\ell \mu}$ distribution using the background model and a simulation of $t\bar{t}$ events. The $t\bar{t}$ samples are generated with different top quark mass values in the range 150--195~GeV/$c^2$, incrementing by steps of up to 0.5 GeV/$c^2$, and the full $M_{\ell \mu}$ spectra are determined by adding the signal and expected background histograms in the ratio shown in Table \ref{tab:sample}. Figure \ref{fig:Control_IM_12Jb} shows the mean value of the $M_{\ell \mu}$ distributions versus the input top quark mass, indicating a linear relationship between the two quantities. Also shown is $<M_{\ell \mu}>=35.6\pm1.1({\rm stat.})$~GeV/$c^{2}$, measured in the data. 
We perform a binned-likelihood fit to the $M_{\ell \mu}$ histogram of the data, in 20 bins between 4--100~GeV/$c^2$, with the binning and range chosen a priori appropriately to the size of the data sample.
The likelihood is defined as:
\begin{eqnarray}
-{\rm ln}L(m_t) & = & -\sum_{i=1}^{\rm N_{bins}} n_{i}^{\rm data}~{\rm ln}\left[\frac{n_i^{\rm TP}(m_t)}{n_{\rm tot}^{\rm TP}}\right],
\end{eqnarray}

\noindent where $n_i^{\rm data}$ and $n_i^{\rm TP}(m_t)$ are the number of entries in each $i$-bin of the data and template histograms respectively, the total number of entries is ${n_{\rm tot}^{\rm TP}}={n_{\rm tot}^{\rm data}}$, and ${n_i^{\rm TP}(m_t)}/{n_{\rm tot}^{\rm TP}}\equiv{{\cal{P}}_i(m_t)}$ is the probability of the $i$-th bin, normalized such that $\sum_i { {\cal{P}}_i}=1$. 
 The background normalization is fixed and its value is varied in the evaluation of the systematic uncertainty. A parabolic function is fit to the values of ${\rm ln}L(m_t)$ derived from each mass template, and the measured top quark mass is determined from the minimum of the likelihood function, while the statistical uncertainty is given by the range corresponding to an increase in the $-{\rm ln}L$ of 0.5 units above the minimum. For each mass point within the full mass range, we generate 5000 pseudoexperiments with the same sample size as that of the data and verify that the fitting procedure is unbiased and that the statistical uncertainty returned by the fits represents the 68\% confidence level. From 248 $t\bar{t}$ candidate events, we measure:
\begin{equation}
 m_t=180.5\pm12.0({\rm stat.})\pm3.6({\rm syst.})~{\rm GeV}/c^2.
\end{equation}

\begin{figure}[h]
\begin{center}
\includegraphics[width=6.5cm]{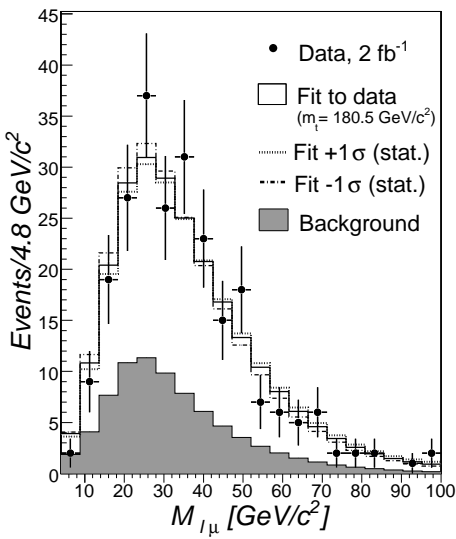}  \\
\end{center}
\caption{{\small The distribution of invariant mass $M_{\ell\mu}$ of the lepton from the $W$ decay and the SLT$\mu$, from a sample of 248 candidate $t\bar{t}$ events with 79.5 background.}}
\label{fig:LKmeasured}
\end{figure}

\noindent Figure \ref{fig:LKmeasured} shows the $M_{\ell \mu}$ distribution of the data, the background, and the templates corresponding to the best fit and the statistical uncertainty.

The sources of systematic uncertainty that affect the measured value of the top quark mass are summarized in Table \ref{tab:syst}. The limited size of the $t\bar{t}$ samples simulated with different values of $m_t$, input to the fitting procedure, yields an uncertainty of $\pm$0.3~GeV/$c^2$. Several components enter the uncertainty on the modeling of the background. The uncertainty on the $W+$ heavy and light flavor normalizations yields an uncertainty of $\pm$0.5~GeV/$c^2$. The uncertainty on the shape of the $W$+jets histogram is evaluated by varying the distribution, to within the statistical accuracy associated with the comparison in the $W$+1,2 jets sample between the data and the background model, and yields an uncertainty of $\pm$1.4~GeV/$c^2$. The normalization of the QCD multijet background contributes $\pm$0.8~GeV/$c^2$. The shape of the QCD multijet distribution accounts for $\pm$0.6~GeV/$c^2$, as determined by replacing the nominal sample with dijet enriched data selected by $I<0.1$ and $\slashed{E}_T<15$~GeV, and by varying the distribution according to its statistical uncertainty. The shift on the measured top quark mass due to the uncertainties on the remaining backgrounds is negligible. 
The total uncertainty from background modeling is $\pm$1.9~GeV/$c^2$. 

Monte Carlo modeling of the signal $M_{\ell \mu}$ distributions includes effects of PDFs, initial-state radiation (ISR), final-state radiation (FSR), and JES. The uncertainty due to the MC modeling of $t\bar{t}$ production and decay, including $b$ fragmentation, is determined by comparing the simulation using {\sc pythia} with that using {\sc herwig} and gives $\Delta m_t=\pm$2.1 GeV/$c^2$. The PDF uncertainty is evaluated by adding in quadrature the contribution of four effects: variations of the PDFs according to the 20 CTEQ eigenvectors \cite{pdf}, the difference between the standard $t\bar{t}$ simulation using the CTEQ5L PDF and one derived using MRST98 \cite{mrst} in the default configuration or with two alternative choices for $\alpha_s$, and the variation of the contribution of gluon fusion in $t\bar{t}$ production between 5 and 20\%.
The overall estimated uncertainty from PDF is $\pm$1.0~GeV/$c^2$. We vary both ISR and FSR simultaneously in the $t\bar{t}$ Monte Carlo simulation, within constraints set by studies of radiation in Drell-Yan events in the data, and assign a systematic uncertainty on $m_t$ of $\pm1.3$~GeV/$c^2$. 

The jet reconstruction is used in this analysis only for the selection of event candidates and therefore the uncertainty on the calibration of the jet energies enters the measurement solely through the event selection, via the jet counting and the $\slashed{E}_T$ requirement. The uncertainty due to the JES is measured by shifting the energies of the jets in $t\bar{t}$ MC simulation by $\pm1\sigma$ of the JES \cite{jetcorr} and results in $\Delta m_t=\pm$0.3 GeV/$c^2$. The uncertainty of $\pm$1\% on the difference between data and simulation of the PL energy and momentum scales gives an uncertainty of $\pm$0.9 GeV/$c^2$. The differences in the data versus simulation for the SLT$\mu$ $p_T$ spectrum depends on the $b$-quark fragmentation modeling and the momentum calibration. In addition to the different fragmentation models in {\sc herwig} versus {\sc pythia}, we consider comparisons of the data with MC simulation of the muon $p_T$ spectra in $B\rightarrow\mu^-D^0X$ \cite{kraus} and $b\bar{b}\rightarrow\mu\mu X$ \cite{paolo} which indicate an uncertainty on the muon $p_T$ of $\sim\pm0.8$\%, corresponding to $\Delta m_t=\pm0.9$~GeV/$c^2$. The uncertainty on the $p_T$ dependence of the SLT$\mu$ tagging efficiency yields a shift on the top quark mass of $\pm$0.2 GeV/$c^2$. Finally, a source of systematic uncertainty is due to the modeling of pile-up events from multiple $p\bar{p}$ interactions and it is estimated to affect the measured mass by $\leq \pm0.5$ GeV/$c^2$. 

\begin{table}[h]
\caption{\small Summary of systematic uncertainties.}
\begin{center}
\begin{tabular}{r}
\begin{tabular}{lc}
\hline
\hline
Source    & $\Delta m_{\rm t}$ [GeV/$c^2$]\\
\hline
MC $t\bar{t}$ samples statistics  &   $\pm$0.3 \\
Background &  $\pm$1.9 \\
$t\bar{t}$ production and decay model   &  $\pm$2.1 \\
Parton distribution functions		    & $\pm$1.0 \\
Initial- and final-state radiation	    & $\pm$1.3 \\
Jet energy scale		    & $\pm$0.3 \\
PL energy/momentum scale  & $\pm$0.9 \\
SLT$\mu$ momentum		 & $\pm$0.9 \\
Pileup	    	    & $\pm$0.5 \\
\hline
Total		  &  $\pm$3.6 \\
\hline
\hline
\end{tabular}
\end{tabular}
\end{center}
\label{tab:syst}
\end{table}

In summary, we have performed the first measurement of the top quark mass in a sample of $t\bar{t}\rightarrow \ell\bar{\nu} b \bar{b}q\bar{q}$ events selected by identifying $b$-jets with a muon candidate from the semileptonic decay of heavy-flavor hadrons. The result, $m_t=180.5\pm12.0({\rm stat.})\pm3.6({\rm syst.})~{\rm GeV}/c^2$, is in agreement with the current world average value of 173.1$\pm$1.3 GeV/$c^2$ \cite{bestmt}, providing a consistency check of the top quark sector with soft muon $b$-tagged events. Our measurement technique exploits the correlation between the parent top quark mass and the invariant mass of the system composed of the lepton from the $W$ decay and the muon from the semileptonic $B$ decay. The uncertainty at present is dominated by the statistical component.
The method has a minimal dependence on the jet energy calibration, making it suitable for averaging the result with those from other techniques, and its dominant systematic uncertainties are likely reducible, e.g. by improving the calibration of the leptons' $p_T$ to better than 1\% with $J/\psi$, $\Upsilon$ and $Z$ resonances, by using improved tuning for the MC modeling of $t\bar{t}$ production and decay, and with high statistics data samples for the background model. 

We thank the Fermilab staff and the technical staffs of the participating institutions for their vital contributions. This work was supported by the U.S. Department of Energy and National Science Foundation; the Italian Istituto Nazionale di Fisica Nucleare; the Ministry of Education, Culture, Sports, Science and Technology of Japan; the Natural Sciences and Engineering Research Council of Canada; the National Science Council of the Republic of China; the Swiss National Science Foundation; the A.P. Sloan Foundation; the Bundesministerium f\"ur Bildung und Forschung, Germany; the Korean Science and Engineering Foundation and the Korean Research Foundation; the Science and Technology Facilities Council and the Royal Society, UK; the Institut National de Physique Nucleaire et Physique des Particules/CNRS; the Russian Foundation for Basic Research; the Ministerio de Ciencia e Innovaci\'{o}n, and Programa Consolider-Ingenio 2010, Spain; the Slovak R\&D Agency; and the Academy of Finland.



\begin{thebibliography}{99}
\bibitem{cdfsys} We use a ($z,\phi, \theta$) coordinate system where the $z$-axis is in the direction of the proton beam, and $\phi$ and $\theta$ are the azimuthal and polar angles respectively.  The pseudo-rapidity is $\eta\equiv-\ln(\tan\frac{\theta}{2})$. Transverse energy and momentum are $E_T\equiv E\sin\theta$ and $p_T\equiv p\sin\theta$ respectively, where $E$ and $p$ are energy and momentum. The missing transverse energy is $\slashed{E}_T \equiv |-\sum_i E_T^i\hat{n}_i|$, where $E_T^i$ is the magnitude of the transverse energy contained in each calorimeter tower $i$ in the region $|\eta|<3.6$, and $\hat{n}_i$ is the direction unit vector of the tower in the plane transverse to the beam direction.
\bibitem{ht} The $H_T$ is defined as the scalar sum of the electron (muon) $E_T$ ($p_T$), $\slashed{E}_T$ and jet $E_T$ for jets with $E_T>8$~GeV and $|\eta|<2.4$.
\end{thebibliography}
\end{document}